\documentclass[a4paper,11pt]{article}

\usepackage{jcappub} 

\usepackage[T1]{fontenc} 
\usepackage{hyperref}
\usepackage{amsfonts}
\usepackage{epsf,graphicx}
\title{Inflaton as a pseudo-Goldstone boson of vacuum energy shift symmetry}

\author[a]{Ja.V.Balitsky}
\author[a,b,1]
{and V.V.Kiselev\note{Corresponding author.}}
\emailAdd{Valery.Kiselev@ihep.ru}

\affiliation[a]{Moscow Institute of Physics and Technology (State University),
Russia, 141701, Moscow Region, Dolgoprudny, Institutsky 9}
\affiliation[b]{Russian State Research Center
Institute for High Energy Physics (National Research Centre Kurchatov
Institute), Russia, 142281, Moscow Region, Protvino, Nauki 1}

\abstract{
The exact symmetry induced by the global shift of energy scale for the
cosmological constant in the action of matter fields is spontaneously broken
by setting the density of real vacuum energy and explicitly broken by the
gravity that means the existence of Nambu--Goldstone boson, which transforms
into the pseudo-Goldstone boson due to the gravity. We identify this boson
with the inflaton in the Einstein frame of action for the fields, while the
breaking is induced by a non-minimal interaction of boson with the scalar
curvature in the Jordan frame. The role of Galilean symmetry of field equation
is emphasized in the procedure of fixing some terms in the bare action.
}

\begin{document}
\maketitle

\section{Introduction and the exact global symmetry}

If the gravity is switched off, then the physics is indifferent with respect
to an absolute value of vacuum energy density $\rho_\Lambda=\Lambda^4$, and
the global shift of scale $\Lambda\mapsto \Lambda+\Lambda_c$ conserves
non-gravitational forces, since it corresponds to the choice of
minimal-energy prescription or its renormalization, which is irrelevant to
the dynamics as constrained by interactions excepting the gravity. Therefore,
the non-gravitational action incorporates the global shift symmetry of
cosmological constant associated with the vacuum energy-density\footnote{See
treatments of the cosmological constant problem in
\cite{Weinberg:1988cp,Weinberg:2000yb}. The role of zero-point modes of
quantum fields for a determination of vacuum density of energy was first
discussed in \cite{Zel'dovich:1968zz}.},
\begin{equation}\label{gl-shift}
    \Lambda\mapsto \Lambda+\Lambda_c.
\end{equation}
However, in the real world the density of vacuum energy is not arbitrary, it
is definite, hence, the global symmetry of non-gravitational forces is
spontaneously broken, since the laws of non-gravitational dynamics are not
changed under this fixing of cosmological constant\footnote{We suggest that
the super-symmetry (SUSY) is inherently related with the gravity, since the
local SUSY gives the super-gravity. See also further discussion in the
text.}. Then, the Goldstone theorem \cite{Goldstone:1961eq,Goldstone:1962es}
states that the global parameter of continuous symmetry transforms into the
local massless field $\phi(x)$, the Nambu--Goldstone boson, interacting with
other fields and conserving the symmetry of action under the global shift
symmetry
\begin{equation}\label{glob-phi}
    \phi(x)\mapsto \phi(x)+f_G\,u,
\end{equation}
where $u$ is the dimensionless parameter of global symmetry, while $f_G$ is
the dimensional constant of energy. The value of $f_G$ is fixed by setting
the canonical expression for the Lagrangian of free field $\phi$: $\mathcal
L_0=\frac12 (\partial_\mu\phi)^2$. The interaction of such Goldstone boson
with the matter can include derivative terms, 
for instance\footnote{We exclude higher derivatives to hold the ordinary
causality, when the maximal order of differentiation in field equations
should not exceed 2.},
\begin{equation}\label{L_G}
    \mathcal L=\frac12 (\partial_\mu\phi)^2+b_e\,\partial_\mu\phi\,j^\mu+
    \frac12\,b_T\,\partial_\mu\phi\,\partial_\nu\phi\,T^{\mu\nu}+\ldots
\end{equation}
where $j^\mu$ is a 
vector current, which should be bilinear in the matter fields at least,
$T^{\mu\nu}$ is a symmetric 
tensor, which should be linear in the matter fields at least, and so on.

The global symmetry (\ref{glob-phi}) generates the Noether current in the
Minkowski space-time
\begin{equation}\label{N-c}
    K^\mu=\frac{\partial \mathcal L}{\partial\partial_\mu\phi}\,
    \frac{\partial\phi}{\partial u}=f_G\left\{
    \partial^\mu\phi+b_e\,j^\mu+b_T\, \partial_\nu\phi\,T^{\mu\nu}
    \right\}.
\end{equation}
This current is conserved in the case of absence of gravity, of course.
Moreover, the conservation law
$
    \partial_\mu K^\mu=0
$ coincides with the field equation for $\phi$. It is important that in the
absence of sources ($j=T=0$) or at $b_e=b_T=0$ there is the solution linear
in coordinates
\begin{equation}\label{linear}
    \phi=f_G\,k_\mu x^\mu+\phi_0,
\end{equation}
which conserves the translational invariance of Minkowski space-time, since
the global translation $x^\mu\mapsto x^\mu+a^\mu$ generates the global shift
$\phi(x)\mapsto \phi(x)+f_G k_\mu a^\mu$, that is the symmetry of action with
the Goldstone boson. Solution (\ref{linear}) remains valid for the case of
non-zero sources, if only the sources are the conserved tensors\footnote{For
the sake of straightforward extension to the case of curved space-time, we
use the general covariant derivative $\nabla_\mu$ instead of partial
derivative of Descartes coordinates in Minkowskian space-time.}
\begin{equation}\label{conservation}
    \nabla_\mu j^\mu=0,\qquad \nabla_\mu T^{\mu\nu}=0.
\end{equation}
Moreover, expression (\ref{linear}) corresponds to the classical background
for the field of massless quanta for $\phi$. In addition, the expansion over
symmetric tensors in (\ref{L_G}) should be
truncated by the shown tensor of second rank, since 
the absolute conservation laws of real word point to that $j^\mu$ is the
electric current, while $T^{\mu\nu}$ is the energy-momentum
tensor\footnote{The energy-momentum tensor  should include the contribution
of field $\phi$ itself, that can produce a recurrence.}, and we have not get
any other conserved symmetric tensors. Then, a non-zero global charge appears
at $k^2>0$, when in the system $k_\mu=(k_0,{\bf 0})$ it can be easily
calculated, while the boson action with (\ref{linear}) is reduced to the
following:
$$
    S_\phi=\int d^4 x\,\mathcal L\Big|_{(\ref{linear})}=\int dt\left\{
    \frac12 f_G^2 k^2(V+b_T E)+b_e f_G\sqrt{k^2} Q
    \right\},
$$
where $E$ is the energy of sources in the whole volume $V$ and $Q$ is their
electric charge in the same volume. For the vacuum $Q=0$, $E=\rho_\Lambda V$
and we get the global shift for the density of vacuum energy:
$\rho_\Lambda\mapsto \rho_\Lambda-\frac12\,f_G^2 k^2(1+b_T\rho_\Lambda)$.

The sense of solution (\ref{linear}) at $k^2>0$ is clear in the inertial
system wherein $k^\mu=(k_0,{\bf 0})$: at the trivial potential the field of
Goldstone boson can move in space homogeneously and isotropically  at a
constant velocity even when the boson interacts with the conserved sources.
Cases of $k^2=0$ and $k^2<0$ break the mentioned homogeneity and isotropy of
space.

For the free Goldstone boson, its energy-momentum tensor under (\ref{linear})
trivially takes the ordinary form
$$
    t_{\mu_\nu}^{(0)}=\left(k_\mu k_\nu-\frac12\,k^2 g_{\mu\nu}\right) f_G^2,
$$
that yields the pressure equal to the energy density $p_0=\rho_0=\frac12\,k^2
f_G^2$ in the system $k^\mu=(k_0,{\bold 0})$. Such the equation of state
gives the ``stiff matter''. The interactions significantly change the value
of parameter $w=p/\rho$, of course.

Let us emphasize that solution (\ref{linear}) can be interpreted as the
introduction of symmetry for the field equations in the form
\begin{equation}\label{G-sym}
    \phi(x)\mapsto \phi(x)+f_G k_\mu x^\mu+\phi_0,
\end{equation}
which has been called the ``galilean'' invariance
\cite{Nicolis:2008in,Deffayet:2009wt}. The corresponding scalar field
possessing the symmetry of (\ref{G-sym}) of its field equations is referred
as the Galileon and it is considered in the gravity modification involving
the scalar degree of freedom, so that it can incorporate some higher
derivative terms of the field in its action, but the field equations remain
the second order equations due to the specified symmetric form of action. In
this way, the symmetry permits three higher derivative terms in the action
with fixed structures \cite{Nicolis:2008in}. In addition, by construction the
Galileon postulates the decoupling limit: it interacts with the matter only,
in the form of contact term of the field with the trace of energy-momentum
tensor. We will see that both features of Galileon, i.e. the higher
derivative action terms and decoupling from the gravity, are beyond the scope
of current study. Thus, we treat the Galilean symmetry of field equation in
the very different aspect, whereas the global shift symmetry is broken due to
the coupling to the gravity, while the higher derivative terms of self-action
are not involved. A study of comparison and connection of our model with the
Galileon, especially with the cosmology due to the Galileon
\cite{Chow:2009fm}, will be given elsewhere. However, we emphasize that the
Galilean symmetry makes our further constructions to be uniquely definite,
since it restricts opportunities in choice of interaction terms.

\section{Couplings and symmetry breaking}

Coupling constants $b_{e,T}$ determine the strengthes of interactions. Note
that global field possesses zero interactions. Moreover, the term with $b_e$
does not influences the equations of motion. Nevertheless, due to the
dimensional analysis it can be represented as the quantity proportional to an
inverse scale of energy,
$$
    b_e=\frac{\kappa}{M},\qquad \kappa=\pm 1,0,
$$
so that $M$ sets the scale of dimensional expansion for the interaction
operators, hence,
$$
    b_T\sim\frac{1}{M^4}.
$$


The gravity does not influence the conservation laws (\ref{conservation}),
but it breaks down the global symmetry (\ref{glob-phi}). The minimal way of
gravitational interaction with the matter conserves the global symmetry of
boson vertices. Therefore, we expect that the gravity has to involve
non-minimal terms in order to break down the global symmetry.

Suggest that the symmetry is restored in the limit of zero scalar curvature,
$\mathcal R\to 0$. In the simplest case the interaction of $\phi$ with the
gravity is induced by the Lagrangian linear in both the curvature and field,
that explicitly breaks the global symmetry\footnote{The linear contact
interaction of boson with the scalar curvature conserves the ``galilean''
symmetry (\ref{G-sym}) of boson field equation. Thus, the galilean symmetry
uniquely fixes the linear interaction.},
\begin{equation}\label{int}
    \mathcal L^\mathrm{int}=-\frac12\,b_\xi\,\phi\,\mathcal R,
\end{equation}
because the shift introduces the finite renormalization of gravitation action
itself\footnote{In units $8\pi\,G=1$, where $G$ is the Newton constant.}:
$$
    \mathcal L_G=-\frac12 \,\mathcal R \mapsto
    -\frac12 \mathcal R\left(1+b_\xi f_G u\right).
$$
Such the breaking could be called ``natural'', since the only effect it
causes is the change of gravity strength.

Switching on the gravity involves a new symmetric interaction in the form
$$
    \mathcal L_{SG}=\frac12\,b_G\,\partial_\mu\phi\,\partial_\nu\phi\left(
    \mathcal R^{\mu\nu}-\frac12\,g^{\mu\nu}\mathcal R\right),
$$
due to Bianchi identity $\nabla_\mu \left(
    \mathcal R^{\mu\nu}-\frac12\,g^{\mu\nu}\mathcal R\right)\equiv 0$.
The coupling constant $b_G$ scales as
$$
    b_G\sim \frac{1}{M^2}.
$$
Taking into account Einstein equations
$$
    \mathcal R^{\mu\nu}-\frac12\,g^{\mu\nu}\mathcal R=T^{\mu\nu},
$$
we expect that the terms of interaction with the energy-momentum tensor and
Ricci tensor should be of the same order, that means $b_T\sim b_G$ and
$M^2\sim 1$ in Planckian units. Thus, the scale $M$ in the effective
expansion over the derivative operators is of the order of Planck mass.

The impact of operator $\mathcal L_{SG}$ on the inflation has been studied in
\cite{Germani:2010hd}, wherein the effect of enhanced gravitational friction
during the slow rolling of inflaton field to a potential minimum was pointed
out in the case of minimal coupling of the field to the scalar curvature. So,
the coupling to the Einstein tensor $R_{\mu\nu}-\frac12 g_{\mu\nu}R$
preserves the inflation in the range of perturbative regime safely beyond the
limit of strong energy scales. Further generalizations of this issue were
considered in \cite{Kobayashi:2011nu,DeFelice:2011hq,VanAcoleyen:2011mj}.
After the general introduction of $\mathcal L_{SG}$ we disregard this term in
our study for the sake of simplicity of demonstration of our basic findings
in order to focus on a field potential induced by gravitational loops that
break the Galilean symmetry of field equations.

Further, as we will show in the next section, gravitational loop-corrections
generate the effective potential starting at the quadratic term\footnote{In
general, a contribution linear in the field gives a global shift associated
with a non-zero value of vacuum  expectation value for the field $\phi$.} in
the mass $m\sim b_\xi$,
\begin{equation}\label{V}
    V=\frac12\, m^2\phi^2+\ldots, 
\end{equation}
so that in the leading order of consideration the potential is quadratic in
the coupling $b_\xi$. In the leading approximation of potential (\ref{V}) and
in the case of no sources ($j=T=0$) or at $b_e=b_T=b_G=0$ the field equation
reads off
\begin{equation}\label{Eq-0}
    \Box\phi=-m^2\phi-\frac12\,b_\xi\,\mathcal R,
\end{equation}
resulting in
\begin{equation}\label{break-K}
    \nabla_\mu K^\mu=-f_G\,m^2\phi-\frac12\, b_\xi\,f_G\mathcal R,
\end{equation}
that represents the ordinary expression for the pseudo-Goldstone boson and it
satisfies the limit of exact global symmetry $\partial_\mu K^\mu\to 0$ at
$b_\xi\to 0$, since $m^2\sim b_\xi^2\to 0$.

\section{Matching to the inflaton}
\subsection{The effective potential induced by gravitational loops}
\begin{figure}[h]
\setlength{\unitlength}{1mm}
\begin{center}
\begin{picture}(110,30)
\put(0,3){\includegraphics[width=30\unitlength]{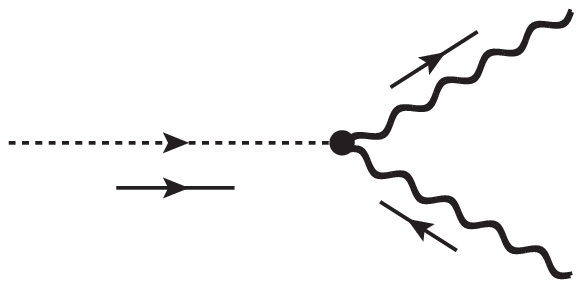}}
\put(60,0){\includegraphics[width=50\unitlength]{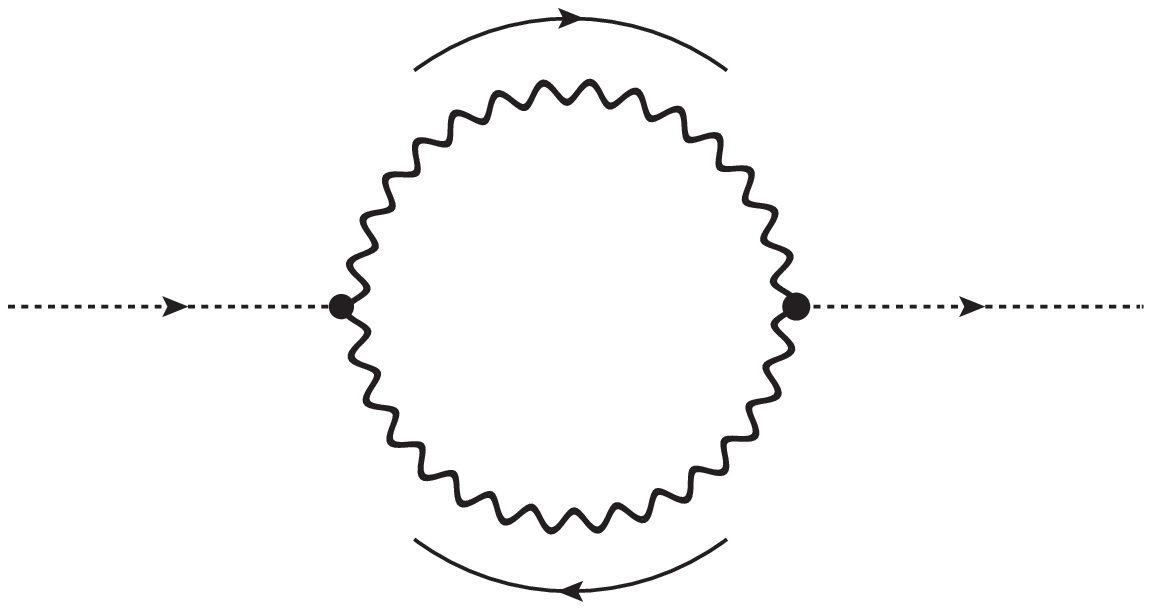}}
\put(4,12){$\phi$}
\put(31,17){$h_{\mu\nu}$}
\put(31,1){$h_{\alpha\beta}$}
\put(8,5){$q$}
\put(17,17){$q+p$}
\put(19,4){$p$}
\put(64,12){$q$}
\put(112,12){$q$}
\put(84,24){$q+p$}
\put(88,-2){$p$}
\end{picture}
\end{center}
  \caption{The graviton coupling to the scalar pseudo-Goldstone boson (left)
  and the graviton loop for the quadratic term of scalar field (right,
  momenta are labeled).}
  \label{f1}
\end{figure}

The contact term of non-minimal interaction generates the
graviton-graviton-scalar vertex $V^{\mu\nu,\alpha\beta}$ in the Minkowskian
background (see figure \ref{f1} (left)),
\begin{equation}\label{vertex}
\begin{array}{rl}
    V^{\mu\nu,\alpha\beta}[p+q,p]=\displaystyle\frac{\xi}{2 m_\mathrm{Pl}}
    \Bigg[& (p^2+q\cdot p+q^2)\eta^{\mu\alpha}\eta^{\nu\beta}+\\ &
    \Big\{(p+q)^\alpha p^\nu\eta^{\mu\beta}+
    (p+q)^\beta p^\mu\eta^{\nu\alpha}-p^\alpha p^\beta \eta^{\mu\nu}-
    \\ &
    (p+q)^\mu (p+q)^\nu\eta^{\alpha\beta}\Big\}\Bigg],
\end{array}
\end{equation}
where we define the graviton as the perturbation of flat metrics
$\eta_{\mu\nu}=\mbox{diag}(1,-1,-1,-1)$ and set $b_\xi=\xi m_\mathrm{Pl} $,
so that $g_{\mu\nu}=\eta_{\mu\nu}+h_{\mu\nu}/m_\mathrm{Pl}$ in Planckian
units in terms of reduced Planck mass $m_\mathrm{Pl}$: $8\pi G
m_\mathrm{Pl}^2=1$.

The graviton propagator is taken in the standard form 
\begin{equation}\label{prop-g}
    \Pi_{\mu\nu,\mu'\nu'}[p]=\frac{i}{p^2+i0}\left(\eta^{\mu\mu'}\eta^{\nu\nu'}+
    \eta^{\mu\nu'}\eta^{\nu\mu'}- \eta^{\mu\nu}\eta^{\mu'\nu'}\right),
\end{equation}
so that after the Wick rotation to the Euclidean momenta, i.e. at
$p^2=-p_E^2$, $d^4p=id^4p_E$, $\eta^{\mu\nu}\mapsto-\delta^{\mu\nu}$, we
regularize the loop diagram in figure \ref{f1} (right)
$$
    i\Gamma^{(2)}[q]=\int\frac{d^4p}{(2\pi)^4}\,iV^{\mu\nu,\alpha\beta}[p+q,p]\,
    \Pi_{\mu\nu,\mu'\nu'}[p+q]\,iV^{\mu'\nu',\alpha'\beta'}[-p-q,-p]\,
    \Pi_{\alpha\beta,\alpha'\beta'}[p]
$$
by introducing the cut-off $\Lambda_c$ via $p_E^2<\Lambda_c^2$ and after a
cumbersome arithmetics we get the quadratic term of effective action
$$
    \Gamma[\phi]=\frac12\,\phi^2 \Gamma^{(2)}[q],
$$
with the effective mass $m$
\begin{equation}\label{G2_0}
    m^2=-\Gamma^{(2)}[0]= c_2 (1-c'_2)\,\xi^2\,\frac{\Lambda_c^4}{m_\mathrm{Pl}^2},
    \qquad c_2=-\frac{39}{64\pi^2},
\end{equation}
and the correction to the kinetic term
\begin{equation}\label{G2_2}
    \frac{\partial^2 \Gamma^{(2)}[0]}{\partial q^\mu\partial q^\nu}=\eta_{\mu\nu}\,
    k_2 (1-k'_2)\,\xi^2\,\frac{\Lambda_c^2}{m_\mathrm{Pl}^2},
    \qquad k_2=\frac{147}{48\pi^2}.
\end{equation}
Here $c_2$ and $k_2$ are the diagrammatic factors, which were calculated,
while $c'_2$ and $k'_2$ are some constants of subtraction in the
renormalization scheme. At $c'_2>1$, we get the stable quadratic potential.

 Generically, the correction to the kinetic term can be of the order of
unit\footnote{One could say about the induced kinetic term, but we have
already got the propagator of scalar field in the calculation of the 1-loop
effective action.}, that means the relation between the coupling constant
$b_\xi$ and the cut-off scale $\Lambda_c$:
\begin{equation}\label{cut}
    \xi^2\sim \frac{m_\mathrm{Pl}^2}{\Lambda_c^2}.
\end{equation}
Therefore, the mass scale of pseudo-Goldstone boson is of the order of loop
cut-off,
\begin{equation}\label{m-cut}
    m\sim \Lambda_c.
\end{equation}
We suggest that $m\ll m_\mathrm{Pl}$, hence, $\xi\gg 1$.

Next, we simplify the consideration by setting the symmetry of effective
potential under the inversion $\phi\leftrightarrow - \phi$. It means that all
of odd terms in $\phi$ are equal to zero strictly under the renormalization
procedure.

Then, we worry about higher powers of $\phi$ in the effective potential. So,
we can analogously evaluate term including $\phi^n$ by introducing the
corresponding cut-off $\Lambda_n$ in the Euclidean space. In this way, we can
easily evaluate the effective potential as
\begin{equation}\label{effect}
    V_\mathrm{eff}=\sum_{n=2j} \frac{c_n}{n!}\,(1-c^\prime_n)\,
    \phi^n \Lambda_n^4\,\frac{\xi^n}{m_\mathrm{Pl}^n}.
\end{equation}
Here, constants $c_n$ are determined by diagrammatic factors, while
$c^\prime_n$ depend on the receipt of renormalization, so that
$|1-c^\prime_n|\sim 1$. The cut-off scales set the energy factors in diagrams
with $n$ gravitons in the loop. Since each contribution determines the
independent quantity in the effective potential, the scales of cut-off are
generically different for each $n$.

The same diagrams generate contributions, which renormalize the kinetic term,
\begin{equation}\label{kinetic}
    K_\mathrm{eff}=(\partial_\mu\phi)^2\sum\limits_{n=2j} \frac{k_n}{n!}\,(1-k^\prime_n)\,
    \phi^{n-2} \Lambda_n^2\,\frac{\xi^n}{m_\mathrm{Pl}^n},
\end{equation}
where $k_n$ and $k^\prime_n$ are diagrammatic and renormalization factors,
respectively. We see that the actual expansion of $K_\mathrm{eff}$ in $\phi$
corresponds to the expansion in powers of inverse Planck mass $m_\mathrm{Pl}$
only if the cut-off scales range as
\begin{equation}\label{cut-n}
        \Lambda_n^2\sim m_\mathrm{Pl}^2\,\frac{1}{\xi^{n}},
\end{equation}
or
\begin{equation}\label{cut-c}
    \Lambda_n^2\sim\Lambda_c^2\,\frac{1}{\xi^{n-2}},\qquad \mbox{at }
    \Lambda_c=\Lambda_2.
\end{equation}
Otherwise the expansion in (\ref{kinetic}) blows up at Planckian values of
field. Therefore, the effective potential takes the form
\begin{equation}\label{effect-c}
    V_\mathrm{eff}=\sum_{n=2j} \frac{c_n}{n!}\,(1-\tilde c^\prime_n)\,
    \frac{\phi^n}{m_\mathrm{Pl}^n} \Lambda_c^2m_\mathrm{Pl}^2\,
    \frac{1}{\xi^{n-2}},
\end{equation}
where $\tilde c^\prime_n$ denote modified renormalization constants in
agreement with (\ref{cut-c}).

Thus, setting the true expansion scale to the Planck mass we see that the
higher orders of potential at $n>2$ are suppressed as $\xi^{2-n}\ll 1$, at
least. We conclude that the leading term of potential is reduced to the
quadratic or mass term, indeed.

\subsection{Transformations of fields}

The breaking term corresponds to the scalar field with the non-minimal
interaction with the gravity and it sets the action in the Jordan frame of
fields in the action. It is well known that the transition to the Einstein
frame takes place by the following conformal substitution of metric:
$$
    g_{\mu\nu}\mapsto \frac{1}{\Omega(\phi)}\,g_{\mu\nu} ,
    \qquad \Omega(\phi)=1+b_\xi\phi,
$$
that results in the action
\begin{equation}\label{Einst}
    S=\int d^4x\,\sqrt{-g}\,\left\{-\frac12\mathcal R
    +\frac12(\partial_\mu\phi)^2\left(\Omega^{-1}+\frac32\,
    \left(\frac{\partial\ln\Omega}{\partial\phi}\right)^2\right)
    -V_E(\phi)\right\},
\end{equation}
with the effective potential
$$
    V_E=\frac{V(\phi)}{\Omega^2(\phi)}.
$$
The kinetic term can take the canonical form under the appropriate
transformation of scalar field $\phi\mapsto\chi$, that deduces the ordinary
gravity with the scalar field. Moreover, the leading quadratic approximation
for the potential produces the flat potential $V_E$ at $\phi\to \infty$.
Then, we arrive to the inflaton relevant to the real evolution of early
Universe \cite{Guth:1980zm,Linde:1981mu,Albrecht:1982wi,Linde:1983gd}.

Indeed, for the effective potential discussed in the previous subsection
$$
    V\mapsto\frac{m^2}{2}\,\phi^2,\qquad \Omega(\phi)\mapsto 1+\xi\,
    \frac{\phi}{m_\mathrm{Pl}},
$$
hence,
$$
    V_E(\phi)\mapsto \frac{m^2 m_\mathrm{Pl}^2}{2(m_\mathrm{Pl}+\xi\phi)^2}\,
    \phi^2.
$$
At large couplings $\xi\gg 1$ and positive values of field\footnote{The
general analysis is presented in \cite{Kallosh:2013tua}.}, the kinetic term
becomes canonical under the substitution
$$
    \Omega(\phi)=1+\xi\,\frac{\phi}{m_\mathrm{Pl}}\approx \mathrm{e}^
    {\frac{\chi}{m_\mathrm{Pl}}\,\frac{\sqrt{6}}{3}},
$$
that yields the potential of inflaton $\chi$,
\begin{equation}\label{inf-pot}
    V_E\approx\frac{m^2m_\mathrm{Pl}^2}{2\,\xi^2}\left(1-\mathrm{e}^
    {-\frac{\chi}{m_\mathrm{Pl}}\,\frac{\sqrt{6}}{3}}\right)^2.
\end{equation}
Since according to (\ref{cut}) and (\ref{m-cut}) $m\sim m_\mathrm{Pl}/\xi$,
the plateau of potential is given by the mass of pseudo-Goldstone boson,
$$
    V_E^\mathrm{inf}=\Lambda_\mathrm{inf}^4 \sim \frac{m_\mathrm{Pl}^4}{\xi^4},
$$
which is consistent with the observations at $\xi\sim 300$ with the inflation
scale $\Lambda_\mathrm{inf}\sim 10^{15-16}$ GeV. The inflaton mass is given
by
\begin{equation}\label{inf-mass}
    m_\mathrm{inf}^2=\frac{2m^2}{3\,\xi^2},
\end{equation}
that yields $m_\mathrm{inf}\sim 10^{13}$ GeV. Both estimates of
$\Lambda_\mathrm{inf}$ and $m_\mathrm{inf}$ are in agreement with the current
phenomenology of Universe inflation.

Thus, at strong coupling $\xi\gg 1$ the model tends to the Starobinsky
inflation \cite{Starobinsky:1980te,Mukhanov:1981xt} due to the attractor in
the space of cosmological parameters
\cite{Kallosh:2013tua,Bezrukov:2007ep,Giudice:2014toa} relevant to the PLANCK
observations \cite{Ade:2013zuv}. Sub-leading terms of potential could break
the attractor conditions, that would be relevant to the recent search for a
valuable amplitude of relict gravitational waves, which were induced by the
Universe inflation, if the foreground polarization generated by the dust is
suppressed in a region of detection. At present, the enforced amplitude of
B-mode of cosmic microwave background radiation detected by BICEP2
\cite{Ade:2014xna} corresponds to such the secondary foreground produced by
the measured dust distribution as the Planck collaboration has reported in
\cite{Adam:2014bub}.

\section{Discussion}

Note that at the inflation stage of Universe expansion the only non-zero
component of $\partial_\mu\phi$ is a constant temporal component
$\partial_0\phi\approx\mbox{const.}$, and the interactions of inflaton with
the energy-momentum tensor, Ricci tensor and electric current are inessential
and they can be neglected, because there are no transformation of energy from
the inflaton to the matter, gravity and charges. In contrast, just after the
inflation $\partial_\mu\phi$ become time-dependent and they constitute
sources for the matter production due to the coupling to $T^{\mu\nu}$ (i.e.
the reheating of early Universe at $b_T\neq 0$) and for the direct generation
of gravitational waves due to the coupling to the Ricci tensor $R^{\mu\nu}$
(the mechanism of gravitational enhancement at $b_G\neq 0$). Then, the
inhomogeneity of $\phi$ caused by its quantum fluctuations copies itself in
the primary inhomogeneity of matter distribution and relict gravitational
waves.

Next, the coupling of inflaton to the electric current, $b_e$, could be
essential for the generation of baryon asymmetry of Universe, if the matter
interactions involve appropriate transitions with photons, like, for
instance, non-symmetric $p\to e^+\pi^0\gamma$ or relevant processes with
neutrinos and charged leptons. In this way, Sakharov's conditions
\cite{Sakharov:1967dj} are satisfied, since the inflaton evolution is not
thermal and it breaks the thermodynamical equilibrium, while the
inhomogeneous inflaton becomes the source for processes with the broken both
CP-invariance and conservation of baryon number in the matter sector of
action. In this respect, a mechanism introducing the scalar gauge field could
be actual as considered in \cite{Guendelman:2014bga}.

Thus, in the offered mechanism of explicit gravitational breaking of global
shift symmetry for the scale of cosmological constant the pseudo-Goldstone
nature of inflaton provides us with the reasonable way to describe the actual
inflation of early Universe.

Finally, we have to mention that SUSY models usually treat the inflaton as
the scalar flat direction of K\"ahler potential that corresponds to the
global shift symmetry, while the superpotential and explicit terms of SUSY
breaking generate a non-trivial potential of inflaton
\cite{Kawasaki:2000yn,Kallosh:2010ug}. In this way, the inflaton is the
pseudo-Goldstone boson, too, but the mechanism is evidently different from
the approach considered in this paper.

In addition, the scenario of ``natural inflation'' \cite{Freese:1990rb}
ascribes the axion as the inflaton. In this way, the axion is considered as
the pseudo-Goldstone boson. Again, we see the treatment, which is very
different in its inherent origin.

In fact, both alternative schemes of global symmetry breaking mentioned above
are in action even in the case, when the gravity is switched off and the
Minkowski space-time is considered, in contrast to our model, which is based
on the crucial role of gravity in the symmetry breaking. We think that the
global shift symmetry of cosmological constant scale is more relevant to the
inflaton dynamics.

The global shift of vacuum energy of matter was also considered in the class
of specific theories \cite{Guendelman:1996jr,Guendelman:1995xe}, wherein the
invariant 4-volume $\sqrt{-g}\,d^4x$ with the metric determinant $g=\det
g_{\mu\nu}$ has been modified to the scalar density $\Phi\,d^4x$ with $\Phi$
depending on the first derivatives of scalar fields so that their equations
of motion result in the conservation law consistent with the global shift of
Lagrangian, that is equivalent to the global shift of cosmological constant.
The presence of standard 4-volume terms in the action in addition to $\Phi$
gives the theory with two invariant measures \cite{Guendelman:1999tb}. In
this way, one can avoid the cosmological constant problem at all
\cite{Guendelman:1996jr}. Further, the involvement of global scale invariance
in the double measure theories allows one to construct realistic models of
inflation \cite{Guendelman:1999qt,Guendelman:1999rj} with the potential
profile similar to the Starobinsky inflation or with the residual
cosmological constant suppressed as in the see-saw mechanism. This way is
effective also in the description of quintessence \cite{Guendelman:2001xy}.
The peculiarities of spontaneous breaking of global symmetry in the double
measure theories were investigated in \cite{Guendelman:1999jb}. Some
additional effects arise if one includes the specific interaction like
$T^{\mu\nu}\nabla_\mu\nabla_\nu \phi$ as was studied in
\cite{Guendelman:2009ck} devoted to a gravitational theory with a dynamical
time, that results in extra conservation laws. As we recognize the two
measure theories represent the different way of considering the shift of
cosmological constant, surely.

We will consider cosmological consequences appearing because of various
renormalization schemes of effective action for the inflaton as induced by
one-loop corrections due to the bare non-minimal contact term, elsewhere.

\acknowledgments This work is supported by Russian Foundation for Basic
Research, grant \# 14-02-00096 (for VVK) and grant \# 15-02-03244 (for JaVB).

\paragraph{Note added.}
After the present paper has been written, authors of \cite{Csaki:2014bua}
have suggested a model of induced gravity with the scale invariance due to
the non-minimal squared coupling of inflaton to the scalar curvature in the
Jordan frame. The model treats the inflaton of Einstein frame as it possesses
the global shift symmetry broken by a correction to the effective potential
losing the scale invariance. Despite of similarity of this approach to our
model we see that the motivations as well as the reasons for the global
symmetry breaking are quite different.

\bibliography{bibGold-f}

\end{document}